\documentclass[prl,twocolumn,calc,epsfig]{revtex4-1}
\usepackage{graphics}
\usepackage{amsfonts}
\usepackage{amsmath}
\usepackage{epsfig}

\newif\ifdraft \drafttrue

\catcode`\ä = \active \catcode`\ö = \active \catcode`\ü = \active
\catcode`\Ä = \active \catcode`\Ö = \active \catcode`\Ü = \active
\catcode`\ß = \active \catcode`\é = \active \catcode`\è = \active
\catcode`\ë = \active \catcode`\ô = \active \catcode`\ê = \active
\catcode`\ø = \active \catcode`\ò = \active \catcode`\í = \active
\catcode`\Ó = \active \catcode`\ú = \active \catcode`\á = \active
\catcode`\ã = \active \catcode`\à = \active
\defä{\"a} \defö{\"o} \defü{\"u} \defÄ{\"A} \defÖ{\"O} \defÜ{\"U} \defß{\ss} \defé{\'{e}}
\defè{\`{e}} \defë{\"{e}} \defô{\^{o}} \defê{\^{e}} \defø{\o} \defò{\`{o}} \defí{\'{i}}
\defÓ{\'{O}} \defú{\'{u}} \defá{\'{a}} \defã{\~{a}}\defà{\`{a}}

\begin{document}

\title{Heavy Solitons in a Fermionic Superfluid}
\author{Tarik Yefsah, Ariel T. Sommer, Mark J.H. Ku, Lawrence W. Cheuk, Wenjie Ji, Waseem S. Bakr, and Martin W. Zwierlein}


\affiliation{MIT-Harvard Center for Ultracold Atoms, Research Laboratory of Electronics, and Department of Physics, Massachusetts Institute of Technology, Cambridge, Massachusetts 02139, USA}

\begin{abstract}

Topological excitations are found throughout nature, in proteins and DNA, as dislocations in crystals, as vortices and solitons in superfluids and superconductors, and generally in the wake of symmetry-breaking phase transitions~\cite{zure96}. In fermionic systems, topological defects may provide bound states for fermions that often play a crucial role for the system's transport properties. Famous examples are Andreev bound states inside vortex cores~\cite{caro64bound}, fractionally charged solitons in relativistic quantum field theory~\cite{Jackiw1976Soliton,Goldstone1981Soliton}, and the spinless charged solitons responsible for the high conductivity of polymers~\cite{Heeger1988SSH}. However, the free motion of topological defects in electronic systems is hindered by pinning at impurities~\cite{Heeger1988SSH,Monceau2012CrystalReview}.
Here we create long-lived solitons in a strongly interacting fermionic superfluid by imprinting a phase step~\cite{burg99soliton,dens00,ande01ring} into the superfluid wavefunction, and directly observe their oscillatory motion in the trapped superfluid. As the interactions are tuned from the regime of Bose-Einstein condensation (BEC) of tightly bound molecules towards the Bardeen-Cooper-Schrieffer (BCS) limit of long-range Cooper pairs~\cite{kett08rivista}, the effective mass of the solitons increases dramatically to more than 200 times their bare mass.
This signals their filling with Andreev states~\cite{Antezza2007FermiSoliton,Scott2011SolitonDynamics,scott2012solitondecay} and strong quantum fluctuations~\cite{Dziarmaga2002depletion,Law2002Fluctuations,Mishmash2009Soliton,Martin2010Soliton,Fran2010Solitonreview,Walczak2012solitonbackreaction}. For the unitary Fermi gas, the mass enhancement is more than fifty times larger than expectations from mean-field Bogoliubov-de Gennes theory~\cite{Scott2011SolitonDynamics,Liao2011Solitons}. Our work paves the way towards the experimental study and control of Andreev bound states in ultracold atomic gases~\cite{Antezza2007FermiSoliton,Lutchyn2011Soliton}. In the presence of spin imbalance, the solitons created here represent one limit of the long sought-after Fulde-Ferrell-Larkin-Ovchinnikov (FFLO) state of mobile Cooper pairs~\cite{Yoshida2007LO,Lutchyn2011Soliton}.

\end{abstract}

\maketitle
The emergence of effective particles from an existing background of matter waves is one of the key concepts in physics, from Landau's quasi-particles to the fractionalised charges in quantum Hall states, the Higgs mode in elementary particle physics and cosmic strings in the early universe~\cite{zure96}.
Superfluids are a paradigmatic form of such quantum matter. They are described by a complex macroscopic wavefunction that is rigid against twists of its phase. The ground state of the superfluid thus has uniform phase, and small perturbations propagate as sound waves. A slower, non-linear excitation - the dark soliton - occurs when the phase is twisted substantially over a short range. In the extreme case of a phase jump by 180 degrees, the wavefunction changes sign and crosses zero at the location of the jump, creating a stationary black soliton. In weakly interacting BECs all bosons reside in the condensate, so the particle density vanishes at a black soliton, and is reduced for a moving dark soliton. Dark solitons in BECs have been created~\cite{burg99soliton,dens00,ande01ring,dutton2001shock,Engels2007fluidflow,Becker2008solitons,Weller2008solitons} and studied extensively theoretically~\cite{Fran2010Solitonreview}. They are well described by solutions to the non-linear Gross-Pitaevskii equation.
In fermionic superfluids, solitons are phase twists in the wavefunction of fermion pairs~\cite{Dzia2005Fermisoliton,Antezza2007FermiSoliton}. In the limit of tight molecular pairing, the molecular condensate is still described by the Gross-Pitaevskii equation and stationary solitons are again devoid of particles. However, in the BCS regime of long-range overlapping Cooper pairs, only a minute fraction of particles near the Fermi surface takes part in pairing, and the reduction of the pair wavefunction at the soliton affects the density only very weakly. Solitons in BCS superfluids should thus be filled with normal fluid. Indeed, the depletion of the pair wavefunction gives rise to Andreev bound states localised at the soliton~\cite{Antezza2007FermiSoliton,Yoshida2007LO}. In three dimensions the soliton is a planar defect, one therefore expects a two-dimensional Fermi gas of Andreev states to be confined to the soliton~\cite{Antezza2007FermiSoliton,Yoshida2007LO}.

In the limit of weak attractive interactions, there exists a direct connection between solitons in a BCS superfluid and those studied in relativistic quantum field theory~\cite{Jackiw1976Soliton} and conducting polymers~\cite{Heeger1988SSH}. Here the size of Andreev bound states becomes much larger than the interparticle spacing, and their energy much smaller than the pairing gap. In this limit, the Bogoliubov-de Gennes equation for the bound state simplifies to the Andreev equation~\cite{Yoshida2007LO}, a Dirac equation where the pair wavefunction $\Delta(z)$ plays the role of a spatially varying mass coupling particles and holes. For vanishing transverse momentum it reads:
\begin{equation}
  \left(-i \hbar v_F \sigma_z \frac{\partial}{\partial z}  + \Delta(z) \sigma_x\right)
  \begin{pmatrix}                                                                                                u_n \\                                                                                                v_n
  \end{pmatrix}                                                                                                = E_n
  \begin{pmatrix}
    u_n\\v_n
  \end{pmatrix}
\end{equation}
Here, $\sigma_{x,z}$ are Pauli matrices, $v_F = \frac{\hbar k_F}{m}$ is the Fermi velocity, $k_F = (3\pi^2 n)^{1/3}$ is the Fermi wavevector related to the total density $n$, $u_n e^{i k_F z}$ and $v_n e^{i k_F z}$ are the quasi-particle amplitudes for particles and holes and $E_n$ is the quasi-particle energy.
When $\Delta(z)$ changes sign, this equation allows for one normalisable zero-energy mode per spin state, localised at the zero crossing of the gap, which famously carries half-integer particle number~\cite{Jackiw1976Soliton} and gives rise to the conductivity of polymers~\cite{Heeger1988SSH}.
A self-consistent solution for the pairing field is~\cite{Heeger1988SSH} $\Delta(z) = \Delta_0 \tanh(z/\xi)$, where $\Delta_0$ is the bulk pairing gap and $\xi = \frac{\hbar v_F}{\Delta_0}$ is the BCS coherence length. This soliton solution is shown in Fig. 1a), along with the density of atoms in the bound state, demonstrating its localised nature.
In a BCS superfluid, the Andreev bound state energy is never strictly zero owing to the non-vanishing kinetic energy cost $\hbar^2/m \xi^2 \sim \Delta_0^2/E_F$ of forming the bound state~\cite{caro64bound,Yoshida2007LO}, where $E_F = \frac{\hbar^2 k_F^2}{2m}$ is the Fermi energy.

If $\Delta(z)$ was restricted to be real, as in the theory of conducting polymers~\cite{Heeger1988SSH}, solitons would be topologically protected. However, at finite temperature solitons can accelerate due to dissipation, and the pairing field acquires an imaginary part. When the soliton reaches a critical velocity, it is expected to decay into sound waves or pair excitations~\cite{spuntarelli2011soliton,scott2012solitondecay}. Apart from this thermodynamic instability, in three dimensional systems the soliton can decay into vortices via the so-called snake instability~\cite{ande01ring,dutton2001shock}. Finally, for strong interactions between fermions, it is a priori not obvious that solitons are stable against quantum fluctuations~\cite{Dziarmaga2002depletion,Law2002Fluctuations,Mishmash2009Soliton,Martin2010Soliton,Fran2010Solitonreview,Walczak2012solitonbackreaction}.

\begin{figure}[h]
    \begin{center}
    \includegraphics[width=3.3in]{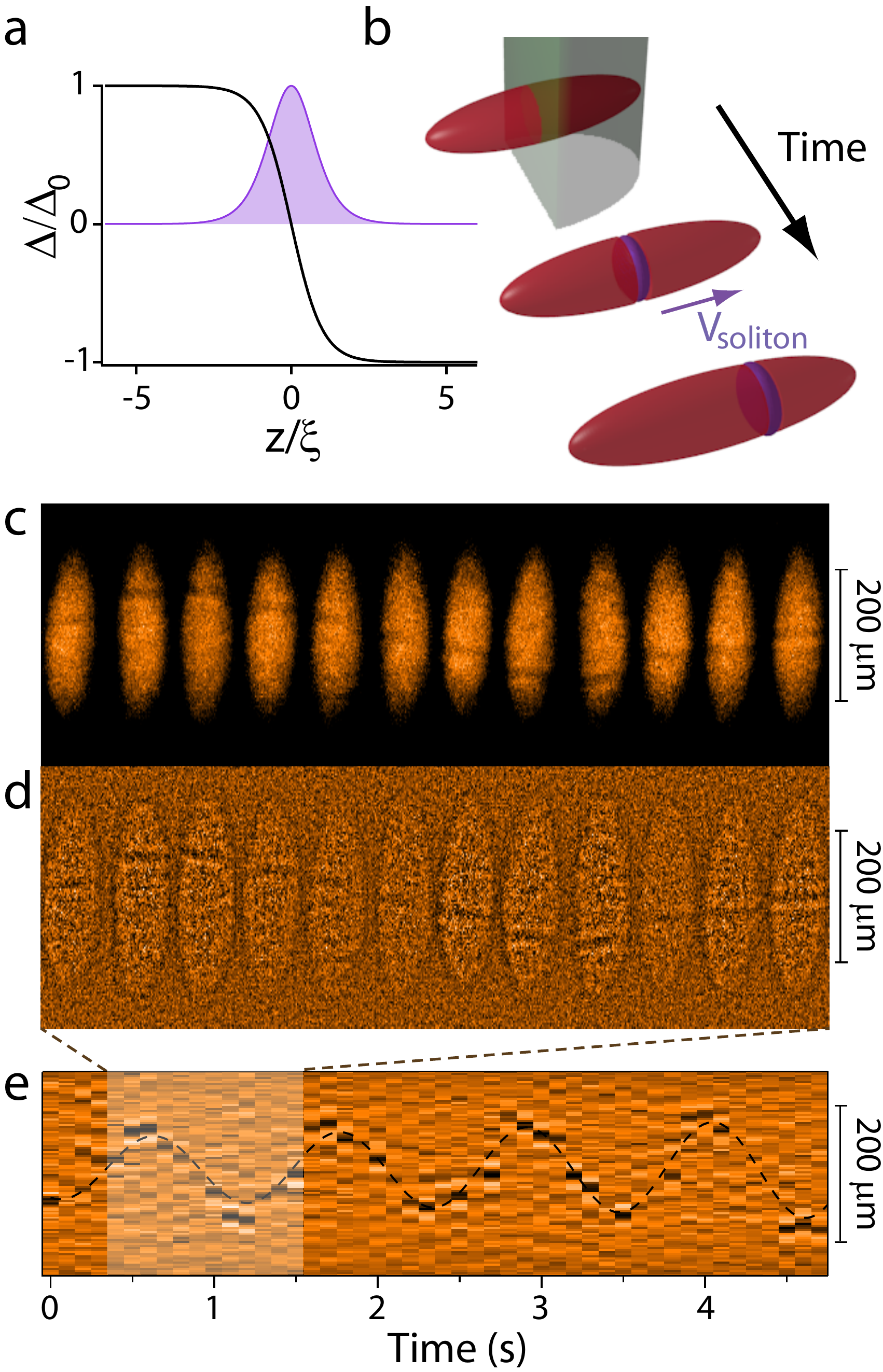}
   \caption[Title]{{\bf Creation and observation of solitons in a fermionic superfluid.} ({\bf a}) Superfluid pair wavefunction $\Delta(z)$ for a stationary soliton, normalised by the bulk pairing gap $\Delta_0$, and localised Andreev bound state density versus position $z$, in units of the BCS coherence length $\xi$. ({\bf b}) Schematic of the experiment. A phase-imprinting laser beam twists the phase of one half of the trapped superfluid by approximately $\pi$. The soliton generally moves at non-zero velocity $v_\mathrm{soliton}$. ({\bf c}) Optical density and ({\bf d}) residuals of atom clouds at 815 G, imaged via the rapid ramp method~\cite{zwie05vort}, showing solitons at various hold times after creation. One period of soliton oscillation is shown. The in-trap aspect ratio was $\lambda=6.5(1)$. ({\bf e}) Radially integrated residuals as a function of time revealing long-lived soliton oscillations. The soliton period is $T_s = 12(2)\,T_z$, much longer than the trapping period of $T_z = 93.76(5)\,\rm ms$, revealing an extreme enhancement of the soliton's relative effective mass $M^*/M$.}
    \label{fig:soliton}
    \end{center}
\end{figure}
Here we create and observe long-lived solitons in a strongly interacting fermionic superfluid of $^6$Li atoms near a Feshbach resonance. Solitons are created via phase imprinting (see Fig. 1b), a technique successfully employed for weakly interacting Bose condensates~\cite{burg99soliton,dens00,Becker2008solitons}. The superfluid containing typically $\sim 2 \times 10^5$ atom pairs is prepared in an elongated trap with cylindrical symmetry (axial and radial trapping period $T_z = 45$ to $210 \,\rm ms$, $T_\perp = 14\,\rm ms$) and tunable aspect ratio $\lambda = T_z/T_\perp$~\cite{kett08rivista}.
A green laser beam far detuned from the atomic resonance is masked to shine on one half of the superfluid. In a time $t$, the applied potential $U$, as experienced by a single fermion, advances the phase of the superfluid order parameter in the exposed region by $\Delta\phi = 2 U t / \hbar$ relative to the unexposed region. The time $t\approx 35\, \mu\rm s$ is experimentally adjusted in order to create one high-contrast soliton.

In the strongly interacting regime, the soliton does not cause a density depletion within our resolution (see Supplemental Material). However, it is tied to a phase twist in the pair wavefunction. As in the case of vortices~\cite{zwie05vort}, the pair wavefunction can be directly observed via a rapid ramp to the BEC-side of the Feshbach resonance. The ramp converts large fermion pairs into tightly bound molecules, empties out the soliton cores and increases the soliton width to the final healing length $\propto 1/\sqrt{n a_f}$, where $a_f$ is the scattering length at the final magnetic field. The rapid ramp followed by time of flight expansion thus enhances the soliton contrast and acts as a magnifying glass (for details see Supplemental Material).

Figure~1~c)-d) reports the observation of solitons in a fermionic superfluid prepared close to the $832\,\rm G$ Feshbach resonance, at $815\,\rm G$, for various hold times following the phase imprint. Here, the interaction parameter at the cloud center is $1/k_F a = 0.30(2)$, where $a$ is the scattering length.
Figure~1c) shows the optical density in absorption images taken after time of flight and the rapid ramp to $\sim 580\,\rm G$, while Fig.~1d) displays residuals obtained by subtracting a smoothened copy of the same absorption image. The optical density contrast is about 10\% (see Supplemental Material). A sequence of radially integrated residuals as a function of time is displayed in Fig.~1~e), demonstrating the soliton to be stable for more than 4 s or $100\,000$ times the microscopic time scale $\hbar/E_F$. This establishes that solitons in fermionic superfluids can exist as stable and long-lived excitations.

The solitons are observed to undergo oscillations in the harmonically trapped superfluid, demonstrating their emergent particle nature. However, their period of oscillation $T_s$ is about one order of magnitude longer than the trapping period $T_z$ for single atoms. This directly indicates an extreme enhancement of their effective mass $M^*$ relative to their bare mass $M = N_s m < 0$, where $|N_s|$ is the number of atoms missing in the soliton. Indeed, the soliton mass times acceleration $M^* \omega_s^2 z$ for periodic motion with frequency $\omega_s$ must be provided by the trapping force experienced by the missing atoms at position $z$, $N_s m \omega_z^2 z$. This yields a direct relation~\cite{Scott2011SolitonDynamics} between the relative effective mass $M^*/M$ and the normalised soliton period $T_s/T_z$:
\begin{equation}
  \frac{M^*}{M} = \left(\frac{T_s}{T_z}\right)^2
\end{equation}
In general, the difference between the effective mass $M^*$ and the bare mass $M$ of the soliton arises from the phase slip $\Delta\phi$ across the soliton, which implies a superfluid backflow~\cite{Scott2011SolitonDynamics}. For the soliton to move, an entire sheet of atoms thus has to flow past it. The difference $M-M^*$ is the mass of that sheet, given by the mass density multiplied by the entire soliton volume. In contrast, the soliton's bare mass $M$ is only due to the mass deficit of $|N_s|$ atoms and can become much smaller than $M^*$ when the soliton is filled. For weakly interacting BECs, where solitons are devoid of particles, the effective mass is still on the same order of the bare mass, $\left(M^*/M\right)_{\rm BEC} = 2$. This leads to an oscillation period that is only $\sqrt{2}$ times longer than $T_z$~\cite{Busch2000soliton,Konotop2004soliton,Becker2008solitons,Weller2008solitons}.
In the BCS limit, where only a minute fraction $\Delta_0/E_F$ of the gas contributes to Cooper pairing, $|N_s| \propto \Delta_0/E_F \propto e^{-\frac{\pi}{2k_F|a|}}$ and thus the soliton's relative effective mass can be expected to become exponentially large.

\begin{figure}
    \begin{center}
    \includegraphics[width=3.3in]{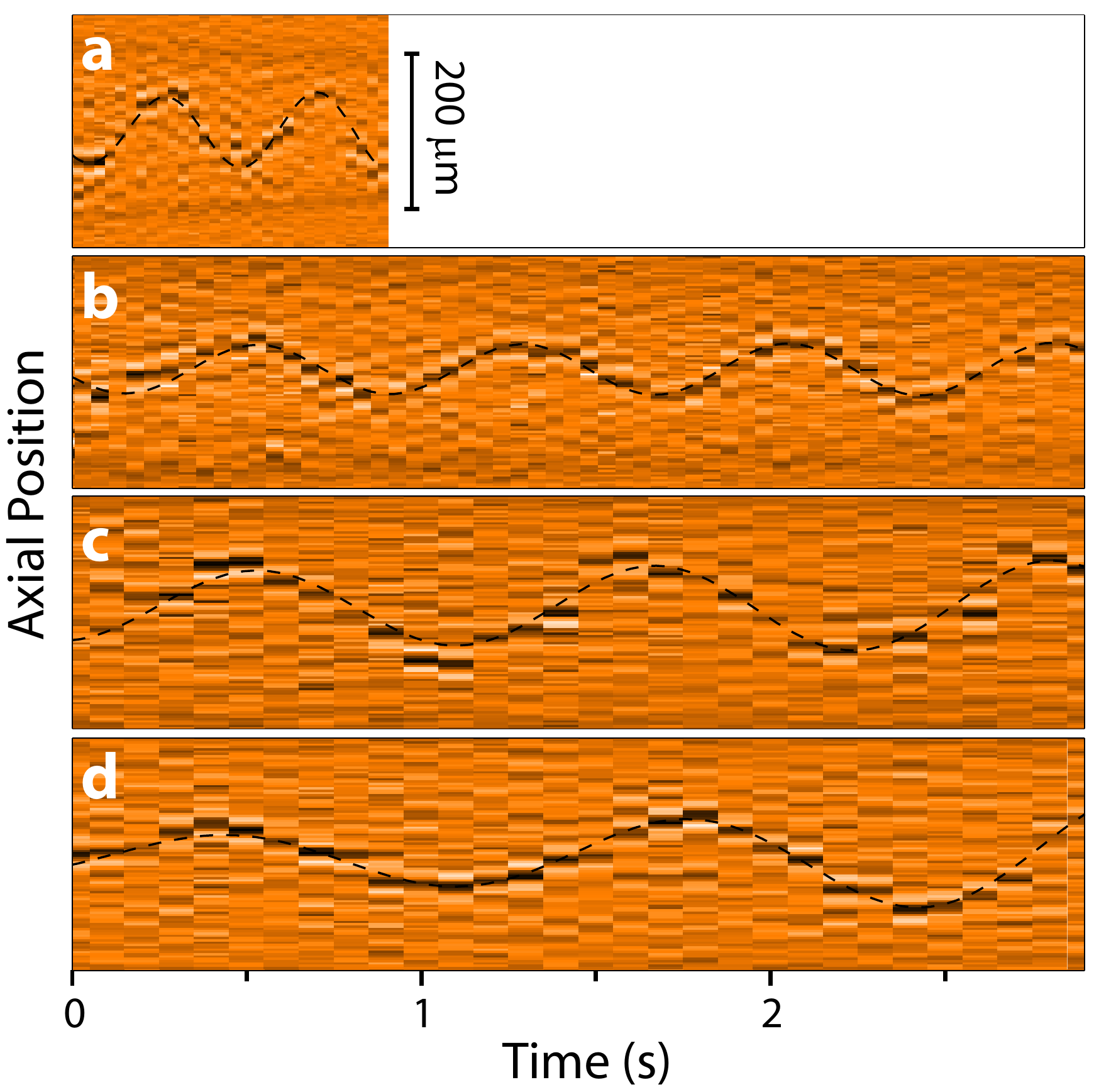}
    \caption[Title]{{\bf Soliton oscillations in the BEC-BCS crossover.} Shown are soliton oscillations in a trapped fermionic superfluid for various magnetic fields $B$ around the Feshbach resonance. The soliton period is observed to dramatically increase as the system is tuned from the BEC regime ({\bf a}) to the Feshbach resonance ({\bf d}).
    The measured period, magnetic field $B$ and interaction parameter at the cloud center $1/k_F a$ were ({\bf a}) $T_s/T_z=4.4(5)$, $700\,\rm G$, $2.6(2)$ ({\bf b}) $T_s/T_z=7.5(9)$, $760\,\rm G$, $1.4(1)$ ({\bf c}) $T_s/T_z=12(2)$, $815\,\rm G$, $0.30(2)$ ({\bf d}) $T_s/T_z=14(2)$, $832\,\rm G$ and $0$. The initial atom number per spin state, its decay rate and Thomas-Fermi radius range from $N_0=1.1\times10^5$, $\tau=1.2(2)\,\rm s$ and  $R_{\rm TF}=135\,\mu\rm m$ at $B=700\,\rm G$ to $N_0=2.3\times10^5$, $\tau=12(1)\,\rm s$ and $R_{\rm TF}=200\,\mu\rm m$ on resonance. The aspect ratio is $\lambda=6.2(7)$. Note that at $B=700\,\rm G$, the superfluid is short lived due to enhanced three-body loss. At $760\,\rm G$ ({\bf b}), the soliton survived for more than $6\,\rm s$, comparable to the lifetime of the superfluid at that field.}
    \label{fig:figure2}
    \end{center}
\end{figure}

Indeed, as shown in Fig. 2, we find that the soliton period, and hence the relative effective mass, increases dramatically as the interactions are tuned from the limit of Bose-Einstein condensation (Fig. 2a) towards the BCS limit. At $700\,\rm G$, where $1/k_F a = 2.6(2)$, the system represents a strongly interacting Bose gas of molecules~\cite{kett08rivista}. The soliton period is $T_s = 4.4(5) T_z$, already three times longer than in the case of a weakly interacting BEC.
At the Feshbach resonance (Fig. 2d), we measure a soliton period of $T_s = 14(2)\, T_z$, corresponding to a relative effective mass of $M^*/M = 200(50)$.
This is more than fifty times larger than the result of mean field Bogoliubov-de Gennes theory in three dimensions~\cite{Scott2011SolitonDynamics,Liao2011Solitons} that predicts $M^*/M = 3$.
Note that the superfluid is fully three-dimensional: on resonance, the chemical potential $\mu \approx 25 \hbar\omega_\perp$, where $\omega_\perp$ is the radial trapping frequency.
Still, for very elongated traps, one expects to reach a universal quasi-1D regime where the tight radial confinement is irrelevant for propagation along the long axis~\cite{Bertaina2010FirstSecondSound}.
This prompted us to study the dependence of the soliton period on the aspect ratio of our trap.

\begin{figure}
    \begin{center}
    \includegraphics[width=3.3in]{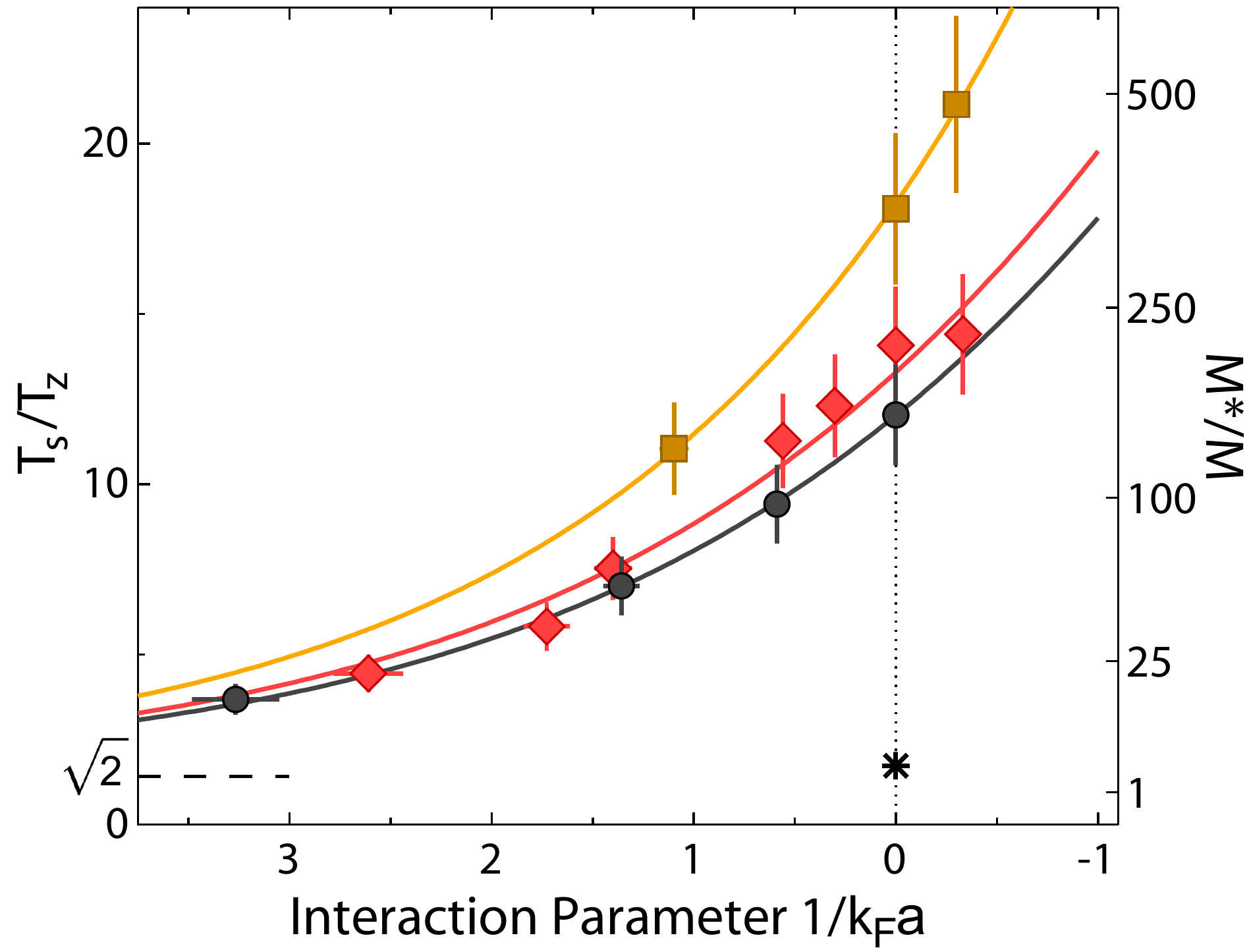}
   \caption[Title]{{\bf Soliton period and effective mass versus interaction strength in the BEC-BCS crossover.} The soliton period is shown as a function of the interaction parameter $1/k_F a$ in the cloud center, for three different trap aspect ratios: $\lambda = 15(1)$ (black circles), $6.2(7)$ (red diamonds) and $3.3(1)$ (orange squares). The soliton period strongly increases from the BEC-regime towards the Feshbach resonance, where $T_s/T_z=12(2)$ for $\lambda = 15(1)$, and to the BCS side. This directly reflects an extreme enhancement of the relative effective mass $M^*/M = T_s^2/T_z^2$, which we attribute to strong quantum fluctuations and the filling with Andreev bound states. The result for a weakly interacting BEC, $T_s/T_z = \sqrt{2}$, is shown as the dashed line. The star marks the mean field prediction~\cite{Scott2011SolitonDynamics} $M^*/M = T_s^2/T_z^2 = 3$.
    }
    \label{fig:figure3}
    \end{center}
\end{figure}

Figure 3 summarizes our measurements for the soliton period and the relative effective mass as a function of the interaction parameter $1/k_F a$ throughout the BEC-BCS crossover, for aspect ratios $\lambda = 3.3$, 6.2 and 15.
The strong increase of $M^*/M$ towards the BCS regime is observed for all trap geometries.
The normalised soliton period $T_s/T_z$ appears to converge to a limiting value for the most elongated trap: The normalised period changes by only 15\% as the aspect ratio is increased by more than a factor of two from 6.2 to 15.
This indicates that the soliton dynamics approach the universal quasi-1D limit.
Even in a much less elongated trap with $\lambda = 3.3(1)$ the soliton period is only slightly increased by about 30\% compared to $\lambda = 6.2$, accompanied by an increased susceptibility of the soliton towards bending or ``snaking''~\cite{ande01ring,dutton2001shock,Fran2010Solitonreview} (see Supplemental Material).

We attribute the large relative effective mass $M^*/M$ in the strongly interacting regime to the filling of the soliton with uncondensed fermion pairs resulting from strong quantum fluctuations. These can also reside inside the soliton~\cite{Law2002Fluctuations,Dziarmaga2002depletion,Mishmash2009Soliton,Martin2010Soliton,Fran2010Solitonreview,Walczak2012solitonbackreaction}, in addition to fermionic Andreev bound states. A substantial filling of the soliton will reduce the number $|N_s|$ of atoms missing inside the soliton, therefore considerably weaken the restoring harmonic force from the trap and strongly increase $M^*/M$. Mean-field theory for the BEC-BCS crossover heavily underestimates the role of quantum fluctuations already on the BEC side, where it predicts a fraction of uncondensed bosons that scales as $n a^3$ instead of the correct $\sqrt{n a^3}$ scaling~\cite{kett08rivista}.
Our experiment thus directly reveals the importance of beyond mean field effects for the dynamics of strongly interacting fermionic superfluids.

\begin{figure*}
    \begin{center}
    \includegraphics[width=5.5in]{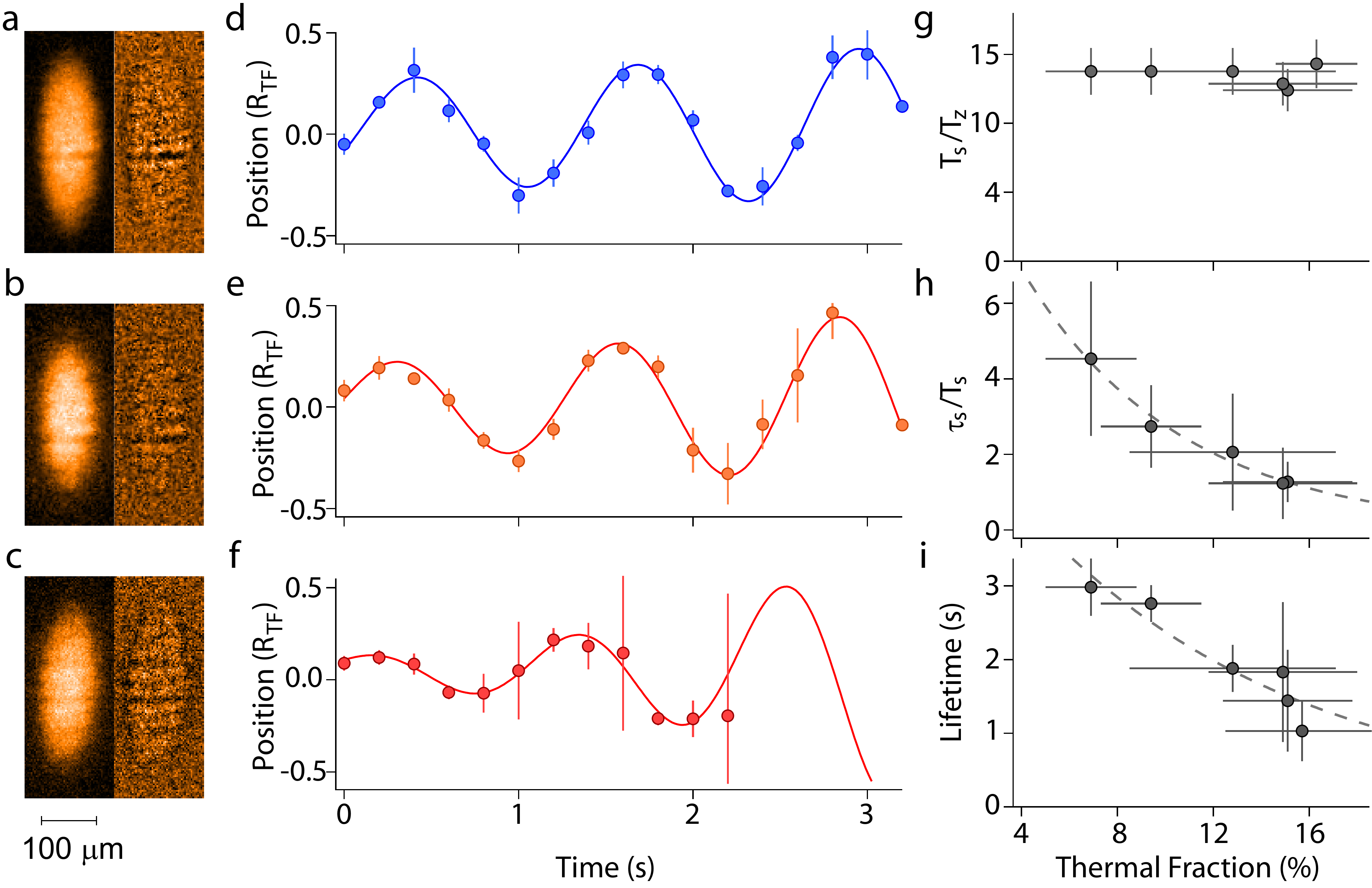}
    \caption[Title]{{\bf Soliton motion in the unitary Fermi gas at various temperatures.}
    {\bf a}-{\bf c} show representative optical densities and residuals of the superfluid after the rapid ramp, and {\bf d}-{\bf f} the trajectory of solitons for thermal fractions $7(2)\%$, $9(2)\%$, and $15(3)\%$, from top to bottom. While at low temperatures, the soliton is the only significant density variation, at higher temperatures transverse stripes appear that we interpret as thermal solitons.
    {\bf g} shows the soliton period versus thermal fraction, demonstrating that the period is insensitive to temperature. The $1/e$ anti-damping time $\tau_s$ ({\bf h}) and the soliton lifetime ({\bf i}) are however strongly dependent on temperature. The soliton lifetime is defined as the time when the probability of observing a soliton decreased to 50\%.}
    \label{fig:figure4}
    \end{center}
\end{figure*}

To demonstrate that the slow soliton oscillations are a truly quantum effect and not due to the finite temperature of our gas, we investigated the soliton motion as a function of temperature for the unitary Fermi gas at the Feshbach resonance (see Fig. 4). The soliton period is found to be insensitive to changes in temperature within the measurement uncertainty (Fig. 4g). The stability of solitons is however strongly affected by the thermal fraction. At low temperatures, the soliton oscillation occurs essentially without energy loss, demonstrating dissipationless flow (Fig. 4d and h). For increasing temperature we observe anti-damping of soliton oscillations (Fig. 4e and h). This is characteristic for a particle with negative mass that can lower its energy by accelerating. The energy loss is likely due to collisions with thermally induced phonons~\cite{Fran2010Solitonreview}. At even higher temperatures, the soliton's position becomes less reproducible and its lifetime is strongly reduced (Fig. 4f and i). Concurrently, we observe increased axial fluctuations in the superfluid (Fig. 4a-c), some of which appear to have comparable contrast to the imprinted soliton. These additional solitons might be ``thermal solitons'', predicted to occur even in equilibrium in weakly interacting Bose condensates~\cite{Karpiuk2012ThermalSoliton}. Similar to vortex-anti-vortex pairs in two dimensions, soliton-anti-soliton pairs can be expected to spontaneously break in 1D and proliferate.
Note that on resonance, the fastest solitons we observe move at the exceedingly slow speed of 0.45 mm/s or 5\% of the (independently measured) speed of sound on resonance. Their sudden disappearance observed e.g. in Fig. 4f can thus not be related to motion close to the Landau critical speed. Instead, their decay might be tied to inelastic collisions with thermal solitons, as soliton collisions have been found to become increasingly inelastic towards the BCS-side in theoretical simulations~\cite{scott2012solitondecay}.
Another possibility for their decay at such low speeds is that the soliton's energy dispersion has a minimum at an unexpectedly small fraction of the critical velocity~\cite{scott2012solitondecay}.

In conclusion, we have created and observed long-lived solitons in a strongly interacting fermionic superfluid. Their period of oscillation and thus their relative effective mass dramatically increases as the interactions are tuned from the BEC-limit of tightly bound molecules towards the BCS limit of long-range Cooper pairs. This signals strong beyond mean field effects, likely due to uncondensed fermion pairs filling the soliton, in addition to purely fermionic Andreev bound states. An exciting prospect is to directly study the Andreev bound states spectroscopically~\cite{Antezza2007FermiSoliton,Lutchyn2011Soliton}. While they are not topologically protected, their lifetime should equal that of the soliton, many seconds or $100\,000$ Fermi times, so that they might become a useful quantum resource.
In the presence of spin imbalance, the soliton represents a limiting case of the long-sought Fulde-Ferrell-Larkin-Ovchinnikov (FFLO) state of moving Cooper pairs~\cite{Yoshida2007LO,Lutchyn2011Soliton}. Indeed, it is energetically favorable for an excess fermion to reside inside a soliton rather than inside the bulk superfluid. While it is difficult to realize the FFLO state in equilibrium, direct engineering of soliton trains might produce a long-lived metastable analog.\\

\textbf{Methods Summary}

{\bf Preparation.} The atomic gas is composed of a balanced mixture of the two lowest hyperfine states of $^6$Li initially prepared at 760 G~\cite{kett08rivista}. The trapping potential results from the combination of a magnetic field curvature providing harmonic confinement in the axial direction and an optical dipole trap (wavelength 1064 nm) providing tighter radial confinement. The axial periods are $T_z=210$~ms, 95~ms and 45~ms, respectively, for the three aspect ratios considered here.

{\bf Phase imprinting.}
A step-like intensity profile is imprinted on a laser beam (wavelength 532 nm, power $\sim$200 mW) by means of an opaque mask. A $3\,\mu$m resolution imaging system projects the intensity distribution at the mask location onto the atoms (the beam waist at the atoms is 60 $\mu$m). A phase twist of $\pi$ corresponds to a pulse time of about $30\,\mu\rm s$, much shorter than the time scale $\hbar/\mu$ associated to a typical chemical potential, on the order of few $100\,\mu\rm s$. The pulse duration is finely adjusted to yield exactly one soliton with high contrast observed after the rapid ramp.
Since we imprint only a phase step but not a density depletion, sound waves must be generated in addition to the soliton~\cite{burg99soliton,dens00}. The sound waves are found to die out in a quarter axial trapping period when they have reached the edge of the atom cloud.
For the data above $760 \,\rm G$ the soliton is created at $760\,\rm G$ and the magnetic field is subsequently ramped (in about 2 $T_z$) to the final magnetic field where the soliton motion is studied. For final magnetic fields below $760 \,\rm G$ the soliton is created at that field. We found that solitons can be created directly at the Feshbach resonance as well.

{\bf Thermometry}
Thermometry from fits of density profiles at the Feshbach resonance to the known equation of state~\cite{ku2012thermodynamics} yielded an upper limit of $T = 0.05 T_F = 10\,\rm nK \approx 3 \hbar \omega_\perp$ for our lowest temperatures. At such low temperatures far below the critical temperature for superfluidity, thermometry via fitting is less sensitive to small changes in temperature~\cite{kett08rivista}. We therefore use the thermal fraction of molecules after a rapid ramp to the BEC-side of the Feshbach resonance as a robust thermometer~\cite{kett08rivista}.

\section*{Supplementary Information}

{\bf Imaging Solitons}

\begin{figure*}[ht]
    \begin{center}
    \includegraphics[width=6.5in]{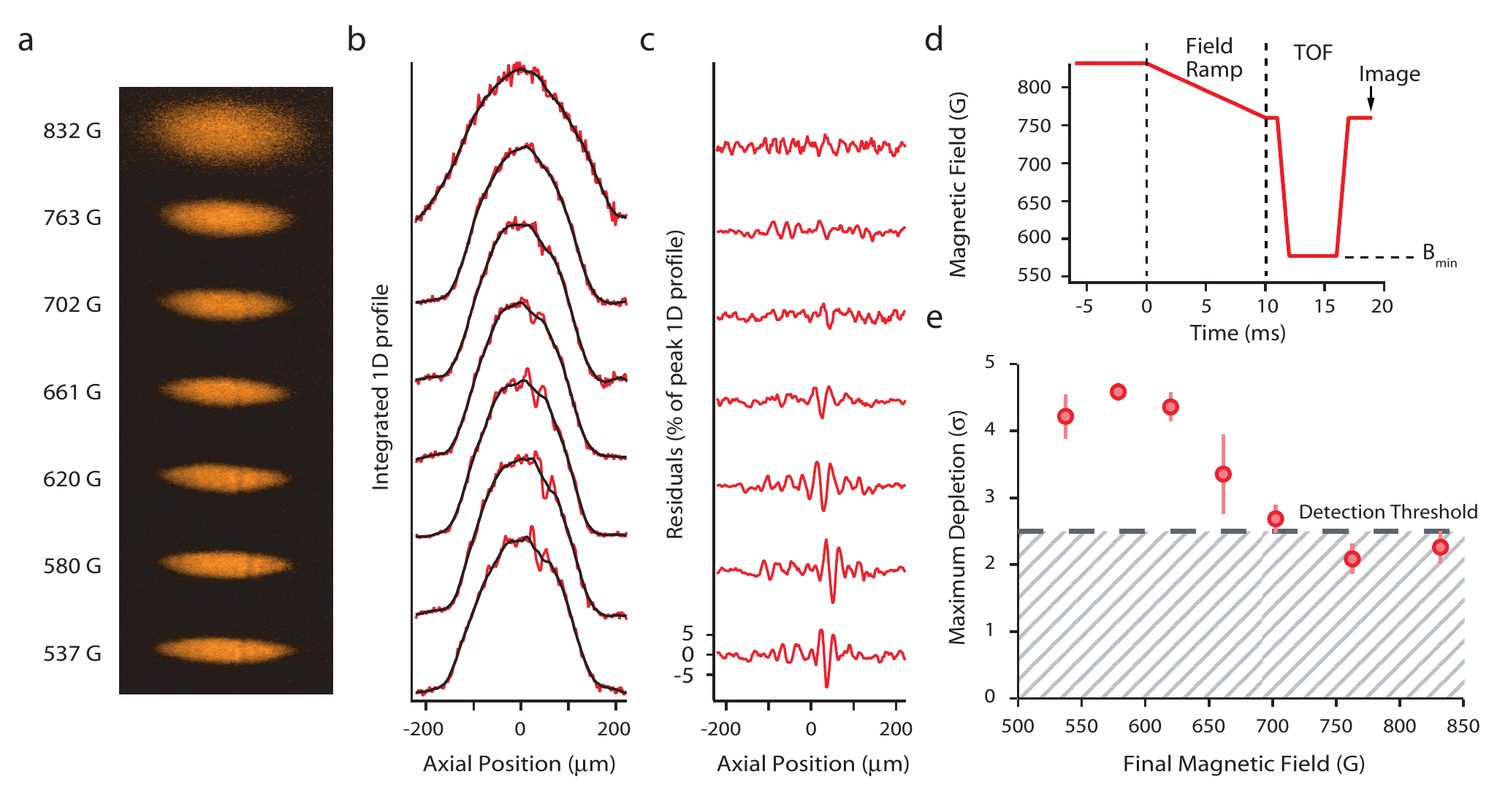}
    \caption[Figure S1]{{\bf Imaging solitons.} {\bf a} Optical density, {\bf b} integrated 1D profiles and {\bf c} corresponding residuals of a fermionic superfluid, prepared at $832 \, \rm G$, after expansion and rapid ramp to various final magnetic fields $B_{\rm min}$. Without any ramp, the superfluid at $832 \, \rm G$, observed after 9~ms time of flight, does not show a clear signature of the soliton. A 10~ms ramp to the BEC-side before expansion at $760 \,\rm G$ reduces interactions but still only reveals a very faint trace of the soliton. For $B_{\rm min} < 700\,\rm G$ the soliton is revealed. {\bf d} Sequence of the magnetic field ramp, indicating time of flight (TOF), final ramp field $B_{\rm min}$ and the imaging pulse at $760 \,\rm G$. {\bf e} The maximum depletion detected in the optical density, in units of the standard deviation $\sigma$ found outside the soliton. The detection threshold of 2.5 $\sigma$ is indicated.
    }
    \label{fig:soliton}
    \end{center}
\end{figure*}

Solitons gradually fill in as the interaction strength is tuned from the BEC-regime to the BCS-regime of the crossover. Indeed, in the BCS regime only a minute fraction $\Delta_0/E_F$ of the gas is Cooper paired, and only this fraction is missing at the soliton's center, where the pair wavefunction $\Delta(z)$ is maximally depleted. The contrast in the particle density thus vanishes. Indeed, absorption images of expanded atom clouds at the Feshbach resonance (Fig.~5a) do not show any observable $(\gtrsim 3\%)$ contrast.
However, the modulus of the pair wavefunction itself can be imaged by a rapid ramp technique similar to what was used for the observation of vortex lattices in the BEC-BCS crossover~\cite{zwie05vort,kett08rivista}. A magnetic field ramp to the weakly interacting BEC-regime turns large fermion pairs into molecules. An absorption image of molecules thus approximately reflects the magnitude of the fermion pair wavefunction before the ramp. In addition, the ramp reduces the interaction strength and thus increases the coherence (healing) length of the superfluid, which increases the soliton contrast and increases their width.

The rapid ramp is illustrated in Fig.~5. Starting for example with a superfluid at the Feshbach resonance, the magnetic field is first quickly ramped over 10~ms to 760 G, on the BEC-side of the Feshbach resonance where interactions are weaker and fermion pairs are more tightly bound. Next, the cloud is released from the trap. After 1~ms, the magnetic field is rapidly ramped over 1~ms to $\sim 580$ G, where interactions are essentially absent and fermion pairs have fully turned into tightly bound molecules. The molecular cloud further expands for 4~ms at $580\,\rm G$, after which the magnetic field is re-ramped over 1~ms to $760 \,\rm G$. After an additional 2~ms of expansion at 760 G, the molecules are imaged via absorption imaging.

Solitons can be identified easily in the absorption images by eye. However, to automize soliton detection we implemented the following method. For each absorption image, we first generate a residual profile by subtracting a smoothened version of the optical density profile from the actual optical density profile. We determine the standard deviation of fluctuations $\sigma$ and identify a depletion in the residual profile as a soliton if its depth is greater than 2.5 $\sigma$.

We have found the rapid ramp technique necessary to reveal solitons in the strongly interacting regime, which is another indication of their strong filling, next to their slow period and enhanced relative effective mass. To show the importance of the rapid ramp, we have varied the final field of the rapid ramp $B_{\rm min}$ between 500 and 832 G. The depth of the maximum depletion, normalized by $\sigma$, is shown in Fig.~5. For ramp fields $B_{\rm min}<650\, \rm G$ solitons are clearly revealed.\\

\begin{figure*}[ht]
    \begin{center}
    \includegraphics[width=3.3in]{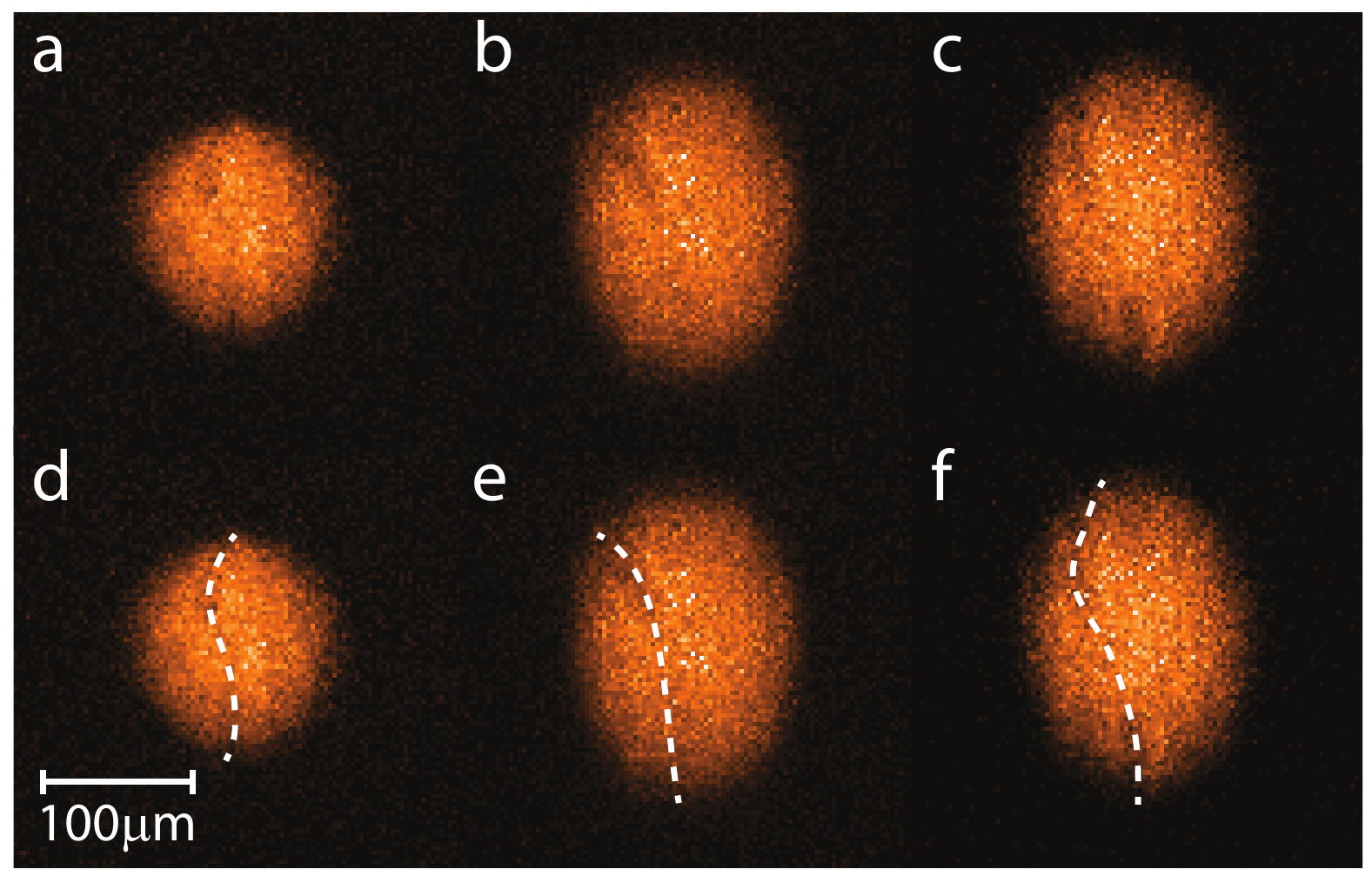}
    \caption[Title]{{\bf Observation of the snake instability in a fermionic superfluid.}
\textbf{a-c} Absorption images exhibiting ``snaking'' solitons in expanding fermionic superfluids.
\textbf{d-f} Same as a-c) but with a white dashed line as guide to the eye.
The superfluids were prepared at $B=760\,\rm G$ in a trap with aspect ratio of $2.9$ (a) and $3.1$ (b-c) with an axial trapping period of $T_z=44\,\rm ms$.
The images were taken $200\,\rm ms$ after the phase imprint of the soliton, using the standard rapid ramp to $B_{\rm min}=580\,\rm G$ and a total time of flight of $9\,\rm ms$.
     }
    \label{fig:soliton}
    \end{center}
\end{figure*}

{\bf Snake Instability}
Planar solitons in three dimensions are not only thermodynamically unstable towards accelerating, but also dynamically unstable towards shape excitations - the so-called ``snake'' instability~\cite{ande01ring,dutton2001shock,Fran2010Solitonreview}.
This excitation of the soliton plane along the radial direction grows exponentially until the soliton decays into vortices. This is suppressed in elongated trap geometries. In weakly interacting BECs, solitons are expected to become dynamically unstable when the chemical potential $\mu$ of the condensate becomes larger than $\approx 2.4 \hbar \omega_\perp$, where $\omega_\perp$ is the radial trapping frequency~\cite{Muryshev1999Soliton}.
The solitons in our strongly interacting fermionic superfluid appear to be much more robust, as they are long-lived even when the chemical potential $\mu \approx 25 \hbar \omega_\perp$.
Still, we were able to observe the snake instability by reducing the trap aspect ratio to about 3. Examples are shown in Fig.~6 where
the depletion revealed by the rapid ramp no longer follows a straight line but rather
a wavy trajectory, characteristic of the snake instability~\cite{ande01ring,dutton2001shock,Fran2010Solitonreview}.

%


{\bf Acknowledgements}
We would like to thank Lev Pitaevskii, Sandro Stringari, Franco Dalfovo, Wilhelm Zwerger and David Huse for fruitful discussions. This work was supported by the NSF, the ARO MURI on Atomtronics, AFOSR PECASE, ONR, a grant from the Army Research Office with funding from the DARPA OLE program and the David and Lucile Packard Foundation.


Correspondence and requests for materials
should be addressed to M.Z.~(email: zwierlein@mit.edu).


\end{document}